\documentclass[universe,preprints,accept,moreauthors,pdftex]{mdpi} 

\firstpage{1} 
\pubvolume{8(12)}
\issuenum{12}
\articlenumber{619}
\pubyear{2022}
\copyrightyear{2022}

\pdfoutput=1

\newcommand{\be}{ \begin{equation} }
\newcommand{\ee}{ \end{equation} }
\newcommand{\beq}{ \begin{eqnarray} }
\newcommand{\eeq}{ \end{eqnarray} }
\newcommand{\x}{\mathrm{x}}
\newcommand{\y}{\mathrm{y}}

\newcommand{\vxd}{{v}_{\mathrm{x}}}
\newcommand{\vxu}{{v}^{\mathrm{x}}}

\newcommand{\vyu}{{v}^{\mathrm{y}}}

\newcommand{\pxu}{{p}^{\mathrm{x}}}
\newcommand{\vxyd}{{v}_{\mathrm{xy}}}

\newcommand{\vyxu}{{v}^{\mathrm{yx}}}

\newcommand{\epsx}{\varepsilon_\mathrm{x}}

\newcommand{\mux}{\tilde{\mu}_\mathrm{x}}

\newcommand{\rx}{\rho_\mathrm{x}}

\newcommand{\n}{\mathrm{n}}
\newcommand{\p}{\mathrm{p}}

\Title{Continuous gravitational wave emission from neutron stars with pinned superfluids in the core}

\Author{Brynmor Haskell$^{1}$\orcidA{}, Marco Antonelli $^{2}$\orcidB{} and Pierre Pizzochero $^{3}$}
\AuthorNames{Brynmor Haskell, Marco Antonelli, Pierre Pizzochero}
\address{%
$^{1}$ \quad Nicolaus Copernicus Astronomical Center of the Polish Academy of Sciences, ul. Bartycka 18, 00-716 Warszawa, Poland\\
$^{2}$ \quad CNRS/in2p3, LPC and ENSICAEN, 14050 Caen, France\\
$^{3}$ \quad INFN and Universit\`a degli Studi di Milano, via Celoria 16, 20133 Milano, Italy %\\
}
\abstract{
We investigate the effect of a pinned superfluid component on the gravitational wave emission of a rotating neutron star. Pinning of superfluid vortices to the flux-tubes in the outer core (where the protons are likely to form a type-II superconductor) is a possible mechanism to sustain long-lived and non-axisymmetric neutron currents in the interior, that break the axial symmetry of the unperturbed hydrostatic configuration. 
We consider pinning-induced perturbations to a stationary corotating configuration, and  determine upper limits on the strength of gravitational wave emission due to the pinning of vortices with a strong toroidal magnetic field of the kind predicted by recent magneto-hydrodynamic simulations of neutron star interiors. 
We estimate the contributions to gravitational wave emission from both the mass and current multipole generated by the pinned vorticity in the outer core, and find that the mass quadrupole can be large enough for gravitational waves to provide the dominant spindown torque in millisecond pulsars.
}

\history{}

\begin{document}
%%%%%%%%%%%%%%%%%%%%%%%%%%%%%%%%%%%%%%%%%%

%%%%%%%%%%%%%%%
%% MAIN TEXT %%
%%%%%%%%%%%%%%%

\section{Introduction}

The detection of Neutron Star (NS) binary inspirals \cite{GW170817} has opened the field of Gravitational Wave (GW) NS astronomy \cite{GW170817multi}, allowing us to place indirect constraints on the internal composition of the star and nuclear parameters of dense matter \cite{GW170817MR,guven_2020PhRvC,Mondal2022PhRvD}. 
Further advances in our understanding of NS physics are expected to come as new detection are made in the next observing run of the LIGO, Virgo and KAGRA network, O4 , see \cite{lasky_gw_astro_2021,Piccinni2022review} for a recent perspective on the topic. 
In particular the instruments are now sensitive enough that one can hope to detect not only transient GW signals from compact object mergers, but also Continuous Wave (CW) signals \cite{Riles22}. These signals are regular, long lived signals, for which galactic rotating NSs are some of the most likely sources \cite{Lasky2015PASA,glamp_gualt_2018, HaskellSchwenzer}. 

One of the main emission mechanisms that has been suggested for CWs from rotating NSs is the presence of a so-called ``mountain'', a quadrupolar deformation that can lead to GW emission at both the stellar rotation frequency and twice the rotation frequency, essentially a GW pulsar. The mountain is sustained either by crustal rigidity \cite{Bildsten98} or by the magnetic field of the star \cite{bonazzola_1996A&A}, see also \cite{Singh2020MNRAS} and references therein. 
In fact, it has been suggested that the observed spin frequencies of accreting NSs in Low-Mass X-ray Binaries may be set by the equilibrium between spin-up torques due to accretion and spin-down torques due to GW emission from mountains created from the asymmetric spread of the accreted material \cite{ucb2000MNRAS,HaskJonesAnd_2006MNRAS}. 
Furthermore, observations of millisecond radio pulsars, which are thought to be old recycled NSs, spun up to millisecond rotation periods ($P$) by accretion in an LMXB, show that there appears to be a cut-off in the allowed parameter space for their evolution in the $P-\dot{P}$ diagram \cite{woan2018ApJL}. 
Remarkably the cutoff can be explained by the presence of a residual deformation, with an ellipticity of $\epsilon\approx 10^{-9}$, which leads to GW emission and causes the systems to evolve out of the excluded region. It has been suggested, see \cite{woan2018ApJL}, that if the core of the star is superconducting, there is a buried magnetic field $B$ of the order of $B\approx 10^{12}$ G. While such a configuration is theoretically possible, it is at the upper limit of what is plausible, given the weak external fields of $B\approx 10^{8}$ G inferred in these systems.

In this paper we shall therefore consider a different mechanism, to understand whether it may play a role in CW emission in rapidly rotating NSs, and whether it could explain the distribution of millisecond pulsars in the $P-\dot{P}$ diagram.
As we have already mentioned the protons in the interior of a mature NS are expected to be superconducting, and the neutrons will also be in a superfluid state. The neutron superfluid rotates by means of quantised vortices, which can interact with the normal component of the star, giving rise to both a dissipative coupling (the so-called mutual friction \cite{Mendell1991,Andersson2006,antonelli_haskell_2020}), but also, if the coupling is strong enough, to pinning to either ions in the crustal lattice \cite{Sevesopin}, or magnetic field flux-tubes in core of the NS \cite{alpar_core_pinning_review}. 

Vortex pinning is at the heart of most pulsar glitch models, which build upon the notion that a pinned superfluid cannot expel vorticity, and thus cannot spin-down, and stores angular momentum. Eventually, once the lag between the superfluid and the normal component becomes too large, hydrodynamical lift forces can overcome the pinning forces, and there is a sudden exchange of angular momentum, i.e. a glitch \cite{ai75,HaskellRev}. The same concept may also be used in reverse to explain anti-glitches in spinning-up pulsars \cite{ducci2015,ray2019ApJ}.

Pinning of a large number of vortices in the crust of a neutron star is generally expected to lead to non-axisymmetric perturbations of the fluid \citep{Ruderman1976ApJ,jones2002_CQG}, and thus possibly to GW emission. While no signal has been detected in post-glitch searches to date  \cite{Glitchsearch1,Glitchsearch2,O3glitch}, some microscopic models of vortex motion in NS crusts predict signals that may be detectable in the near future \cite{jones2002_CQG,bennett_2010MNRAS, WarMel2012MNRAS, Melatos_douglass_2015ApJ}.

While most glitch models focused on pinning to the crustal lattice \cite{ai75}, recent observations of large glitch activity in the the Vela pulsar, combined with theoretical calculations of the entrainment parameter in the crust \cite{Chamel12,Chamel2017} have shown that the moment of inertia of the crust is not enough to explain the amount of spin-down reversed by glitches over time \cite{andersson_notenough,chamel_notenough}, see also \cite{Montoli2021Univ} for a revision of the original theoretical argument and inclusion of statistical uncertainty on glitch activity estimates.
This points towards a picture in which part of the extra angular momentum transferred in a glitch should be stored in the core, a conclusion consistent also with simplified models of maximum glitch sizes in combination with the average activity \cite{montoli2020MNRAS}, post-glitch relaxation \cite{Haskell2018MNRASletter}, and analysis of the spin-up in 2016 Vela glitch \cite{montoli2020A&A}.

In particular the presence of a stronger toroidal field in the interior of the star, which leads to regions of strong pinning of vortices to superconducting flux-tubes (see \cite{alpar_core_pinning_review,sourie_pinning_core_2020} and references therein), could explain some glitch features \cite{guerci_2014ApJ,sourie_vela_2020} like glitch overshoots \cite{pizzochero_over,montoli2020A&A}. 
From the theoretical point of view the situation is unclear, as models have been constructed with toroidal components of the field that can be up to two orders of magnitude stronger than the inferred exterior dipole \cite{Ciolfi2014AN}, but the inclusion of superconductivity in the models generally leads to the expulsion of the toroidal flux to the crust of the star \cite{lander_2013PhRvL,sur2021PASA}, restricting the amount of angular momentum that could be stored in the outer core to power a glitch. 

While further GRMHD modelling is likely to shed light on the internal configuration of the field (see e.g. \cite{sur2022MNRAS} for recent models), in this paper we take as a starting point the observational clues that vortex pinning to flux-tubes may be present in the core and interact with each other \cite{Ruderman1998}, and investigate the consequences for GW emission. 
In particular, we consider models with strong toroidal fields, such as those proposed by \cite{ciolfi2013MNRAS}, in which the magnetic axis is possibly inclined with respect to the rotation axis. This breaks the axisymmetry of the problem, and leads to non-axisymmetric regions of strong pinning, in which vortices may accumulate and lead to velocity perturbations and a mass current quadrupole, resulting in CW emission. 
It has been shown by \cite{jones_2010MNRAS} that in the presence of a pinned superfluid, a deformed rotating NS may emit GWs at both the rotation frequency and twice the rotation frequency, hence we will consider both emission in the $l=2,m=1$ and $l=2,m=2$ harmonic in our model.
 
The paper is organised as follows. In Section 2 we introduce the basic two-fluid formalism and the hydrodynamic force associated to the presence of pinned vorticity. The relation between vortex pinning and the magnetic field configuration is outlined in Section 3, together with the geometry of the pinning region. Section 4 is devoted to discussing stationary perturbations around a global equilibrium solution of the equations introduced in Section 1.
The formalism needed to extract the gravitational wave emission associated to these stationary perturbations is given in Section 5. In Section 6 we present our numerical estimates for the continuous gravitational wave emission due a triaxial deformation of the star induced by the presence of pinning regions. We summarize our findings in Section 7. 

\section{Setting the stage: hydrodynamic perturbations}

As a first step we will present the formalism to describe the perturbed structure of a superfluid NS, modelled in terms of two components, a `proton' component (in reality a charge-neutral fluid of protons and electrons) and a superfluid `neutron' component (see \cite{LivingReviews} and \cite{HasSed} for a review of the formalism).

In a rotating frame of constant angular velocity vector $\mathbf{\Omega}$, and in the limit of slow chemical reactions, the equations of motion for the two species are:
%\cite{Khomenko2019PhRvD}
%
\begin{align}
& D_\x \, \pxu_i + \epsx \vyxu_j \, \nabla_i \, \vxd^{j}+
 \nabla_i \Phi_\mathrm{x}
+ 2 \, \epsilon_{ijk} \Omega^j \vxd^{k}
=
 f_{i}^{\mathrm x} \, 
\label{Euler}
 \\
 & \partial_t \rx +\bigtriangledown ^j  (\rho_x v_{j}^{x})=0 
\label{continuity}
\end{align}
where the chemical index $\x =n$ refers to the neutrons, while $\x =p$ is used for the ``proton'' component; the index $\y$ is used for the species other than $\x$, e.g. $\y=p$ if $\x=n$. The vectors $\vxd^{i}$ and $\pxu_i$ are, respectively, the kinematic velocity field and the momenta per unit mass  of the $\x$ species,  $\vyxu_i=\vyu_i-\vxu_i$ is a shorthand for the relative velocity between the two components (also referred to as `lag'') and $D_\x=\partial_t +  \vxd^{j} \nabla_j$. 
In the left-hand side, $ f_{i}^{\mathrm x} $ is the mutual friction force per unit mass, which is an hydrodynamic coupling between the two species that arises when many quantized vortices are present in the local fluid element \cite{Andersson2006,antonelli_haskell_2020}. 
The effective potential $\Phi_\mathrm{x}$ accounts for gravity (via the gravitational potential $\Phi$), pressure gradients (via the chemical potential per unit mass $\tilde{\mu}_x$) and centrifugal force:
\beq
\Phi_\mathrm{x} = \Phi+\tilde{\mu}_x- |\mathbf{\Omega} \times \mathbf{r}|^2/2   \, .
\eeq
The two thermodynamic quantities that appear in equation \eqref{Euler}, namely $\mux$ and $\epsx$, represent the specific chemical potential and the entrainment parameter of each species (see \cite{thermo} for a formal discussion of superfluid thermodynamics).
In \eqref{Euler}, the non-dissipative entrainment coupling between the two fluids enters also into the definition of the two specific momenta $\pxu_i$, see e.g. \cite{Chamel2017,HasSed},
\beq
\pxu_i \, = \, \vxu_i +  \epsx \, \vyxu_i \, = \, (1-\epsx)  \vxu_i + \epsx \vyu_i   
\quad \text{for} \quad \y\neq \x \, .
\label{Momentum}
\eeq 
All these quantities are local macroscopic variables: the momentum $p^n_i$ is the average momentum per unit mass of a superfluid element that contains several quantized vortex lines, see e.g. \cite{antonelli_haskell_2020}. Since the number of vortices is a physical quantity independent on the frame of reference, to calculate the mutual friction we have to use the the absolute vorticity, namely the macroscopic smooth field defined as
\beq
\omega^i \, = \, \epsilon^{ijk} \, \nabla_j \, p^n_k  \, +  \, 2 \, \Omega^i  \, .
\label{w}
\eeq
The unit vector $\hat{\boldsymbol{\omega}} = \boldsymbol{\omega} /|\boldsymbol \omega|$ defines a preferred direction, along which the quantized vortex lines are (on average) locally aligned. When this is the case, the vortex-mediated mutual friction reads \cite{Andersson2006,antonelli_haskell_2020}
\beq
f_{i}^\x 
\, = \, 
\frac{\rho_n}{\rho_x} \left(
\mathcal{B}^{'} \, \epsilon_{ijk} \omega^j \vxyd^{k} 
\, + \,  
\mathcal{B} \, \epsilon_{iab} \hat{\omega}^a  \epsilon_{blm} {\omega}^l \vxyd^{m}
\right)  \, ,
\label{eq1}
\eeq
where $\mathcal{B}'$ and $\mathcal{B} $ are two dimensionless parameters that encode the physics of the complex dissipation processes at the vortex scale, see  \cite{Graber2018,celora2020MNRAS,antonelli_haskell_2020} and references therein. 
As shown in kinetic simulations of a fluid elements with many vortices, the parameters  $\mathcal{B} '$ and $\mathcal{B} $ depend on the state of motion of the vortex ensemble \cite{antonelli_haskell_2020}: for perfect pinning (as we will consider in the following) we are in the dissipationless limit, where $\mathcal{B} '=1$ and $\mathcal{B} =0$, namely
\beq
\mathbf{f}^n 
\, = \, 
  \boldsymbol{\omega}  \times \mathbf{v}_{np} 
 \qquad \qquad
\mathbf{f}^p
\, = \, 
-\frac{\rho_n}{\rho_p} \, 
 \boldsymbol{\omega}  \times \mathbf{v}_{np}
 \, .
\label{pinning}
\eeq

\section{Magnetic field and pinning}
\label{setup}

Below temperatures of approximately $10^9$ K, the protons in the core of a NS are expected to form a type-II superconductor \cite{HasSed,Chamel2017}, in which the magnetic field is organised in flux-tubes (see \cite{Charbonneau2007PhRvC} for the possibility of having a type-I superconductor). 
Superfluid neutron vortices are magnetised as a consequence of the entrainment effect, and there is thus an energy penalty associated with them cutting through a magnetic flux-tube \cite{Ruderman1998,alpar_core_pinning_review}. This interaction effectively leads to vortices pinning to flux-tubes, below a relative velocity (for a core magnetic field strength $B$)
\begin{equation}
    w\approx 1.5\times 10^4\, {\left(\frac{B}{10^{12}\,\mbox{G}}\right)}^{1/2}  \mbox{cm/s}
    \label{pinf}
\end{equation}
above which hydrodynamical lift forces will unpin the vortices \citep{LinkCore}, see also \citep{sourie_pinning_core_2020}. 
This estimate is valid for the so-called S-wave superfluid, namely where Cooper pairs of neutrons are in the spin singlet state \citep{Sedrakian_Clark_review}. If, on the other hand, neutron pairing is in the P-wave channel (i.e., neutrons form pairs in the triplet state), then the pinning strength is expected to be suppressed, lowering the estimate of the critical lag in \eqref{pinf}, see \citep{Leinson2020MNRAS}. 

As can be seen from \eqref{pinf} the strength of the pinning therefore depends on the large scale structure of the magnetic field, essentially because this dictates the flux-tube density. The equilibrium structure of the magnetic field in the interior of a NS is still an open problem \cite{Pili2017MNRAS}, nevertheless several models have been developed, and there is a general consensus that to be stable, the field must have a twisted-torus configuration, in which a strong interior toroidal component stabilises the poloidal component of the field which stretches outside the star \cite{Ciolfi2014AN,Castillo2017MNRAS,sur2020MNRAS}. 
The relative strength of the toroidal field is particularly interesting, as while many models require this component to be somewhat weaker (roughly an order of magnitude) than the poloidal component \cite{lander_2013PhRvL,sur2021PASA}, there are several models that predict stronger toroidal components in the interior \cite{ciolfi2013MNRAS,pili2014MNRAS}, which would lead to regions of strong vortex-flux-tube pinning \cite{alpar_core_pinning_review}. 
From the observational side, there is some indication that a pinned superfluid must exist in the core of at least the Vela pulsar, to explain the large fraction of spin-down that is inverted by its glitches over time (the activity, see the discussion in \cite{Montoli2021Univ}), but also to explain the difference in post-glitch relaxation with respect to the Crab pulsar \cite{Haskell2018MNRASletter}, and the behaviour of moderately active pulsars displaying glitches of large size like PSR J1341-6220 \cite{montoli2020MNRAS}. 

\begin{figure}%[H]
\centering
\includegraphics[width=13 cm]{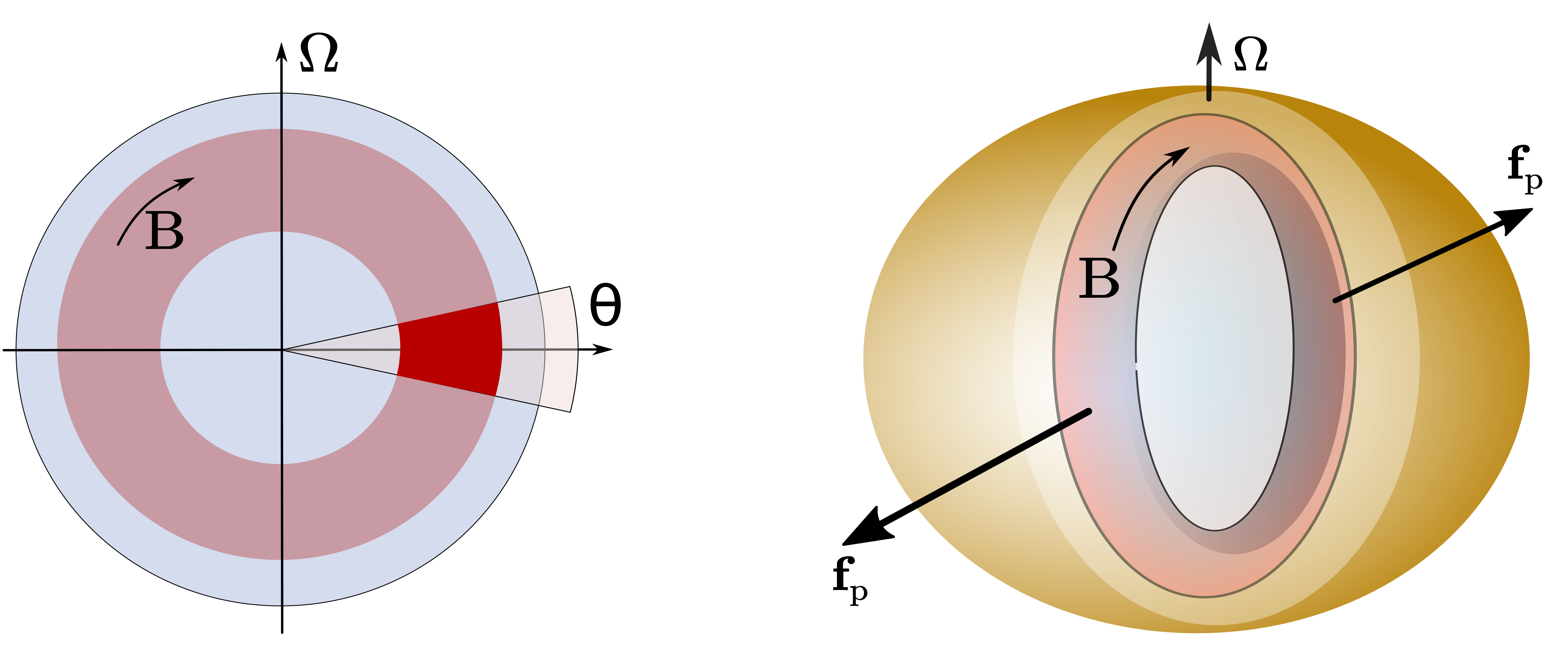}
\caption{
    Left: lateral view of our simplified configuration, in which the neutron star has been sliced along the vertical plane containing the toroidal magnetic field region (its width goes from about $ 0.4R_*$ to $ 0.8R_*$). The two axis are the rotation axis of the neutron star, aligned with the angular velocity $\Omega$, and the $(\theta=\pi/2, \phi=0$), while the dark red region corresponds to the strong pinning region, as indicated in the text. 
    Right: cartoon of the triaxial deformation induced by the presence of pinned vortices in the two, diametrically opposed, strong pinning regions. Because of perfect pining, the force $\mathbf{f}_p$ in \eqref{pinning} is directed outwards when the superfluid spins faster than the normal component. 
    Both sketches are not to scale. 
}\label{fig:1}
\end{figure}   

In order to calculate the strongest GW signal we consider the case of an orthogonal rotator, in which the magnetic field axis is perpendicular to the rotation axis. This may be expected to be the case in a newly born NS, but not throughout the entire population of pulsars. It is nevertheless a useful upper limit. In this case, vortices (which we assume to be straight on a mesoscopic scale) are aligned to toroidal flux-tubes, and will be strongly pinned when they overlap with the closed line region, and essentially free outside. 
In the following we shall consider an approximation to the models of \cite{ciolfi2013MNRAS}, which maximises the strength of the toroidal component in the core of the star: the closed field-line region, in which the toroidal field resides, stretches from about $r_{in}=0.4R_*$ to $r_{out}=0.8R_*$, and is symmetric around the equatorial plane. Therefore, for definiteness, we will assume that the vortices are pinned in a region between $r_{in}$ and $r_{out}$, $\theta_{min}=\pi/2-\arccos{(r_{in}/r_{out})}$ and $\theta_{max}=\pi/2+\arccos{(r_{in}/r_{out})}$, and $\phi_{min}=2\pi-0.08\pi$ and $\phi_{max}=0.08\pi$, as sketched in Figure \ref{fig:1}.

\section{Perturbation equations for the pinned configuration}
\label{sec:pinningregion}

The general equations governing the perturbations of the hydrodynamic model defined by \eqref{Euler}, \eqref{continuity} and \eqref{eq1} are derived in \cite{Khomenko2019PhRvD}. Here, to estimate the effect of a pinned superfluid on the stellar structure and internal flow, we can consider a less general setting, in which the small velocity lag given by the pinning of the neutron vortices to the superconducting flux-tubes is treated as a small perturbation of a background configuration in which the two fluids are corotating.

We consider a constant density background model where neutrons and protons are locked together, while in the strong pinning region sketched in Fig. \ref{fig:1} we have $\mathcal{B}^{'}=1$, and there is a rotational velocity lag given by \eqref{pinf}, such that one has $\delta v_p^i=0$ and  
\begin{equation}
   \delta v^\phi_n 
   = w \, \Theta(r-r_{min})\Theta(r_{max}-r)\Theta(\theta-\theta_{min})\Theta(\theta_{max}-\theta)
   \Theta(\phi-\phi_{min})\Theta(\phi_{max}-\phi) \, ,
   \label{velpert} 
\end{equation}
where $\Theta$ is the unit step function. 
In principle, only the part of this region in which the S-wave superconductor and S-wave superfluid coexist should be considered. Therefore, the exact extent of the pinning region sketched in Fig. \ref{fig:1}, as well as the strength of pinning \citep{alpar_core_pinning_review}, both depend on the exact stratification of the star, hence on the equation of state for nuclear matter (and its consistent dependence of the pairing gaps on density, see e.g. \citealt{Sedrakian_Clark_review}) and on the mass of the star. 
In fact, nuclear models that favour P-wave pairing of neutrons in the outer core are expected to greatly shrink the pinning region to only the lowest-density part of the outer core, as pinning between flux-tubes and topological defects in the P-wave superfluid is probably ineffective \citep{Leinson2020MNRAS}. 
However, since our goal is to provide an upper-limit estimate, we will consider the full region in Fig. \ref{fig:1} as effective, employing there the estimate for pinning between fluxtubes and S-wave superfluid vortices discussed by \cite{LinkCore}, see also \cite{sourie_pinning_core_2020}. In a more realistic scenario both the extent of the pinning region, as well as the pinning strength may be reduced\footnote{
    Apart from the uncertainties on the extent of the S-wave gap for superfluidity, further reduction in the pinning strength may come by the fact that the mutual orientation between a vortex and the fluxtubes is expected to fluctuate locally, giving rise to a decrease of the effective pinning force, similarly to what happens for vortices that are randomly oriented with respect to the principal axis of the Coulomb lattice in the inner crust \citep{Sevesopin}.
}.

Since our goal is to provide a first estimate, we consider also incompressible perturbations ($\delta\rho=0$) and the Cowling approximation ($\delta\Phi=0$).
Following \cite{AndGlampHask_2009}, we can write the perturbed equilibrium equations for the fluids in terms of co-moving and counter-moving variables, which gives for the perturbations of the density and velocity:
\beq
\rho\delta v_j &=& \rho_n\delta v^n_j + \rho_\p \delta v_j^p=\rho_n\delta v^n_j 
\\
\delta w_j &=& \delta v_j^p - \delta v_j^n= - \delta v_j^n \, ,
\eeq
 valid for constant density incompressible fluids with  $\delta v^i_p=0$.
 The continuity equations take the form (assuming time independent perturbations):
 \beq
 \nabla_j(\rho\delta v^j)= \nabla_j(\rho_\n\delta v_n^j)&=&0
 \label{contTOT}
 \\
 \frac{1}{\rho}\nabla_j[x_p(1-x_p)\rho\delta w^j]+\delta v_j\nabla^j x_p&=&0
 \label{contDIF}
\eeq
where we remind the reader that we are considering constant density $\rho$ and proton fraction $x_\p$, and incompressible fluids ($\delta\rho=0$), as well as the Cowling approximation ($\delta\Phi=0$).
Finally one has the momentum equations for the co-moving perturbations:
\be
\nabla^i\delta\Phi+\frac{\nabla^i\delta p}{\rho}+2\epsilon^{ijk}\Omega_j w_k=0\label{eulerTOT}
\ee
with $\nabla^i p=\rho_\n\nabla^i\tilde{\mu}_\n+\rho_\p\nabla^i\tilde{\mu}_\p$, and for the counter-moving perturbations:
\be
\nabla^i(\delta\mu_\p-\delta\mu_\n)+2\bar{\mathcal{B}}^{'}\epsilon_{ijk}\Omega^j \delta w^k-\bar{\mathcal{B}}\epsilon_{ijk}\hat{\Omega}^k\epsilon^{klm}\Omega_l\delta w_m=0 \label{eulerDIFF}
\ee
where $\bar{\mathcal{B}}=\mathcal{B}/x_p$ and $\bar{\mathcal{B}}^{'}=1-\mathcal{B}^{'}/x_p$. We can see immediately that for a velocity perturbation of the form in \eqref{velpert} the total and countermoving continuity equations \eqref{contTOT} and \eqref{contDIF} are satisfied in all the star, except locally at the edge of the pinning domain, due to the sharp boundaries defined by the Heaviside step functions  in \eqref{velpert}. 
They are, however, satisfied in an average sense if one integrates over a volume containing the boundary. Finally from (\ref{eulerTOT}) and (\ref{eulerDIFF}) one can calculate the pressure perturbations $\delta p$ and chemical imbalance $\delta\mu_\p-\delta\mu_\n$ due to pinning.

\section{Continuous wave emission}

In NS interiors large non axisymmetric mass and velocity perturbations can couple to the background metric and lead to the emission of GWs: the emission can be either continuous (e.g., in the case of long-lived ``mountains'') or burst-like, as expected from pulsar glitch events \cite{vE_melatos_2008CQG,bennett_2010MNRAS,sidery_phenomenology_2010,WarMel2012MNRAS}. 
The GW strain $h_{TT}$ is generally described in terms of a multipole expansion, where the leading order is the $l=2$ multipole (the quadrupole), such that:
\begin{equation}
   h^{TT}_{ij}=\frac{G}{c^4 r} \sum_{l=2}^\infty \sum_{j=-l}^l \left( \frac{d^l}{(dt)^l} I_{lm} (t-r) T^{E2}_{lm,ij} + \frac{d^l}{(dt)^l} S_{lm} (t-r) T^{B2}_{lm,ij}\right)
\end{equation}
where $T^{E2}$ and $T^{B2}$ are electric and magnetic tensor spherical harmonics and $I$ and $S$ are generalized mass and current multipole moments \cite{thorne_review_80}, such that for the leading order ($l=2$) terms are
\begin{eqnarray}
I_{2m}&=&\frac{16\sqrt{3}\pi}{15}\int \tau_{00} Y_{2m}^{*}r^2 d^{3}x \\
S_{2m}&=&\frac{32\sqrt{2}\pi}{15}\int (-\tau_{0j}) Y_{22,j}^{B*}r^2 d^{3}x 
\end{eqnarray}
respectively, where $\tau$ is the stress-energy tensor, $Y$ are scalar and $Y^{B}$ magnetic vector spherical harmonics.
As a first step we will consider an incompressible fluid, which naturally leads to a vanishing mass quadrupole under our working hypotheses. This is a common approximation, as this is generally not thought to be the dominant contribution for superfluid flows \cite{WarMel2012MNRAS}. In the following we will however go beyond this approximation and show that, in fact, it can be the dominant contributions for the model we consider. 
For slow motion we have for the mass quadrupole:
\begin{equation}
\tau^{00}\approx \rho
\end{equation}
and for the current multipoles:
\begin{equation}
\tau^{0j}\approx\frac{1}{c}(\rho_\n v^j_\n+\rho_\p v^j_\p)\, .
\end{equation}
If we assume a rigidly rotating background in which the two fluids (the superfluid and the normal component) are corotating, the only contribution to the current quadrupole is due to he perturbations in the superfluid neutron velocity in the pinned regions. We can therefore write:
\begin{equation}
S_{2m}=\frac{32\pi}{15\sqrt{3}c}\int Y^*_{2m}r^2 x^i \epsilon_{ijk}\nabla^j(\rho_\n\delta v_\n^k) d^{3}x 
\label{multipole} \end{equation}
The resulting gravitational wave luminosity takes the form:
\begin{equation}
\frac{dE_{gw}}{dt}=\frac{G}{32\pi c^5}\sum_m |\dddot{S}_{2m}|^2+|\dddot{I}_{2m}|^2
\end{equation}
Clearly, for our uniform density and incompressible model, $\delta \rho = 0 $ and $\tau^{00}= \rho$ is spherically symmetric, so that the mass quadrupole terms vanish, and the current multipole emission dominates. Nevertheless, in a realistic neutron star compressibility will lead to density perturbations associated with the velocity perturbations associated with the pinning. In order to estimate these, in the following we will make the approximation that
\be
\delta\rho\approx\frac{\delta p}{c_s^2}\label{deltarho}
\ee
where $c_s$ is the (constant) speed of sound, and we obtain $\delta p$ by solving the total Euler equation in (\ref{eulerTOT}), leading to
\be
\delta \rho = -2\Omega w \rho r c_s^{-2} \sin\theta\, .
\ee
This is an order of magnitude estimate, as the problem we are solving is incompressible.

\section{Results}

We evaluate the integrals in \eqref{multipole} for the setup described in section \ref{setup}, assuming therefore a perturbation in the neutron velocity only in the strong pinning region. Given the equatorial symmetry, the $l=2, m=0$ and $l=2, m=2$ components of the current multipole vanish, and the leading order contribution is the $l=2, m=1$ multipole (similarly to the case examined by \cite{WarMel2012MNRAS} for vortex avalanches). However also in this case the symmetry we have assumed in the equatorial plane for the two portions of the pinning region lead to the $m=1$ contribution vanishing, so that the $l=2$ current multipole vanishes, i.e. $S_{2m}=0$.
In realistic MHD simulations, however, the field configuration is significantly more complex than in our simple setup \cite{sur2020MNRAS}, as required also by pulse profile modelling with the NICER X-ray telescope \citep{Nicer2019,Raaijmakers2021ApJ}. 
If the strong pinning regions are not symmetric in the equatorial plane, there will thus be small contributions to the $S_{21}$ multipole, the strength of which will depend on the degree of asymmetry.
Let us also consider, however, the contribution to the the mass quadrupole $I_{22}$ due to the density perturbations in the pinning region, as estimated from (\ref{deltarho}). 
In this case we obtain
\be
I_{22}\approx - 2\times 10^{38} \left(\frac{B}{10^{12} G} \right)^{1/2}\left(\frac{\nu}{100\,\mbox{Hz}}\right)\,\mbox{ g cm$^2$}\label{pinnedq}
\ee
where $\nu$ is the rotation frequency of the star, and the GWs are emitted at $2\nu$. We have evaluated our expression for a representative constant density of $\rho=4\times 10^{14}$ g/cm$^3$, corresponding to the pinning region of the outer core. Note however that in our model the pinning region extends throughout the outer core, in a range of densities from $\rho\approx 2\times 10^{14}$ g/cm$^3$ to $\rho\approx 10^{15}$ g/cm$^3$.
Our estimate is approximate, and likely to be an upper limit, due to the extrapolation from the incompressible case. It is clear, nevertheless, that the mechanism should be considered in more detail, as it could explain the residual ellipticity inferred by \cite{woan2018ApJL} in the millisecond radio pulsars.
Furthermore, the scaling with the strength of the magnetic field $B$ is weaker than in the case of standard deformations due to the Lorentz force, in which case one would have \cite{lander_2013PhRvL}
\be
I^{L}_{22}\approx 3\times 10^{37} \left(\frac{B}{10^{12} G} \right)\left(\frac{H_{c1}}{10^{16} G} \right)\,\mbox{ g cm$^2$}
\ee
where $H_{c1}$ is the lower critical field of the type-II proton superconductor  \citep{pb_jones1975,Easson1977PhRvD,lander_2013PhRvL}. We see that the weaker scaling with the surface field strength $B$ of the result for a pinned superfluid in (\ref{pinnedq}) means that even in the case of a weaker buried field of the order of $B\approx 10^9$ G, as is the expected to be the case in the millisecond radio pulsar population \cite{haskell_priymak_2015MNRAS}, one would still expect $\epsilon\approx 5\times 10^{-9}$, and possibly a weaker emission in the $Y_{21}$ multipole, depending on the asymmetry of the field structure.

\section{Summary and conclusions}

In this paper we have evaluated the strength of the mass and current quadrupoles for a rotating superfluid NS with a type-II superconducting core, in which vortices pin to magnetic flux tubes in the region where the toroidal field is stronger. We considered the case of an orthogonal rotator, in which the magnetic field geometry is a twisted torus, and the extent of the closed field line region in which the toroidal component is located is defined in such a way as to approximate the results of \cite{ciolfi2013MNRAS}.
This defines two opposite regions of the star in which neutron vortices are strongly pinned, and a large velocity lag can be sustained between the neutron superfluid and the `normal' component of the star. We evaluate the current quadrupole associated with this velocity lag, and the associated mass quadrupole due to the density perturbation in the pinned region, which necessarily arises to ensure mass continuity.

We find that, due to the symmetries we have assumed, the current quadrupole (the $l=2, m=1$ multipole) vanishes, but the mass quadrupole (the $l=m=2$ multipole) does not, and is in fact sizable, and leads to an ellipticity of the order of
\begin{equation}
    \epsilon\approx - 10^{-9} \left(\frac{B}{10^{8} G} \right)^{1/2}\left(\frac{\nu}{100\,\mbox{Hz}}\right)\, .
\end{equation}

This estimate should be considered an upper limit, as we have assumed incompressible perturbations, an idealized geometry for the magnetic field and the most favourable case of an orthogonal rotator. The quadrupole we obtain is, however, in line with the order of magnitude estimates of \citet{jones2002_CQG}. %mainly due to the pinning force acting in a large and high density region in the core.  
It is thus an interesting result, as it describes a physical mechanism may provide the ellipticity needed to explain the observed cutoff in the $P-\dot{P}$ diagram \cite{woan2018ApJL}. 
Furthermore, the scaling with $\sqrt{B}$, unlike the deformation caused by the Lorentz force which scales as $B$ \cite{lander_2013PhRvL}, is interesting: it implies that also for weaker fields (of the order of $B\approx 10^{8}-10^{9}$ G, more in line with what is expected for millisecond pulsars) one may still have $\epsilon\approx 10^{-9}$, without invoking buried fields of the order of $B\approx 10^{12}$ G, which make it difficult to reconcile theoretical models with the lack of observation of strong GW driven spin-down in the millisecond radio pulsars \cite{haskell_priymak_2015MNRAS}.

Being an upper limit, our estimate is, admittedly, based on some simplifying working assumptions, as  discussed in Sec. \ref{sec:pinningregion}:
our simple treatment does not immediately makes clear which is the role of the specific properties of nuclear matter like the equation of state and the values, nature and density dependence of the pairing gap. In fact, all these effects are indirectly encoded within the estimate of the pinning strength in \eqref{pinf} and the extent of the pinning region.  
Therefore, it will be interesting to further corroborate our upper limit on the GW emission with more refined compressible models of the kind derived in \citep{giliberti2020MNRAS}, in which it is possible to test the role of the equation of state and finite compressibility of nuclear matter.
 
%%%%%%%%%%%%%%%%%%%%%%%%%%%%%%%%%%%%%%%%%%
\vspace{6pt} 
%%%%%%%%%%%%%%%%%%%%%%%%%%%%%%%%%%%%%%%%%%

\acknowledgments{
    Partial support comes from PHAROS, COST Action CA16214. 
    B.H. acknowledges support from the Polish National Science Centre grants SONATA BIS 2015/18/E/ST9/00577 and OPUS 2019/33/B/ST9/00942.
}

\reftitle{References}

\externalbibliography{yes}
\bibliography{biblio}

%%%%%%%%%%%%%%%%%%%%%%%%%%%%%%%%%%%%%%%%%%
\end{document}